\documentclass[aps,pre,twocolumn,amssymb,amsfonts,floatfix,showpacs]{revtex4}

\usepackage{graphicx}
\usepackage{dcolumn}
\usepackage{bm}
\usepackage{mathbbol}

\usepackage{amsmath,amssymb,amsthm}
\usepackage{times}
\usepackage{graphicx}
\usepackage{bm}
\usepackage[dvips]{color}

\begin{document}

\title{Localization and the effects of symmetries in the thermalization 
       properties of one-dimensional quantum systems}

\author{Lea F. Santos}
\email{lsantos2@yu.edu}
\affiliation{Department of Physics, Yeshiva University, New York, NY 10016, USA}
\author{Marcos Rigol}
\email{mrigol@physics.georgetown.edu}
\affiliation{Department of Physics, Georgetown University, Washington, DC 20057, USA}

\begin{abstract}
We study how the proximity to an integrable point or to localization as one approaches 
the atomic limit, as well as the mixing of symmetries in the chaotic domain, may
affect the onset of thermalization in finite one-dimensional
systems. We consider systems of hard-core bosons at half-filling with nearest neighbor
hopping and interaction, and next-nearest neighbor interaction. The latter breaks 
integrability and induces a ground-state superfluid to insulator transition.
By full exact diagonalization, we study chaos indicators and few-body observables. 
We show that when different symmetry sectors are mixed, chaos indicators associated 
with the eigenvectors, contrary to those related to the eigenvalues, capture the onset 
of chaos. The results for the complexity of the eigenvectors and for the expectation 
values of few-body observables confirm the validity of the eigenstate thermalization 
hypothesis in the chaotic regime, and therefore the occurrence of thermalization. 
We also study the properties of the off-diagonal matrix elements of few-body observables
in relation to the transition from integrability to chaos and from chaos to localization.
\end{abstract}

\maketitle

\section{Introduction}

The usual approach to the study of systems with complex energy spectra, such as nuclei,
atoms, molecules, and quantum dots, is via random matrices. These are matrices filled 
with random numbers whose sole restriction is to satisfy the symmetries of the system 
under investigation \cite{MehtaBook,HaakeBook,Guhr1998,ReichlBook}. Time-reversal 
invariant systems with rotational symmetry, for instance, 
are described by the so-called Gaussian orthogonal 
ensembles (GOEs), which consist of ensembles of real symmetric random matrices. 
The range of applicability of random matrix theory (RMT) was further extended after 
the connection with classical chaos became established. It was verified that the spectra 
of quantum systems that behave chaotically in the classical limit show the same fluctuation 
properties obtained with ensembles of random matrices. This observation was stated in the 
form of a conjecture \cite{Bohigas1984} and initiated the field of quantum chaos.

The most commonly used quantities to identify the onset of quantum chaos are based on 
the eigenvalues of the Hamiltonian describing the system under investigation, however 
the structures of the eigenvectors have also a very important role 
\cite{Izrailev1989,Izrailev1990,ZelevinskyRep1996,Flambaum1997,IzrailevREVIEW}.
The eigenvectors of a system whose classical counterpart is chaotic are expected to be 
maximally delocalized. According to Berry's conjecture \cite{Berry1977,BerryProceed},
the eigenfunctions become superpositions of plane waves with random phases and 
Gaussian random amplitudes. 

Berry's conjecture has been connected to the problem of thermalization in 
isolated quantum systems \cite{Deutsch1991,Srednicki1994,Srednicki1996,ZelevinskyRep1996,rigol08STATc}.
As stated in Ref.~\cite{Srednicki1994}, ``a bounded, isolated quantum system of many particles 
in a specific initial state will approach thermal equilibrium if the energy eigenfunctions 
which are superposed to form that state obey Berry's conjecture''. 
It is possible to show that such eigenstates lead to 
the appropriate (Maxwell-Boltzmann, Bose-Einstein, or Fermi-Dirac) distribution for the momentum 
of the particles in the system~\cite{Srednicki1994}.
In this scenario, eigenstate 
expectation values (EEVs) do not fluctuate between eigenstates that are close in energy and 
hence they coincide with the microcanonical average. The latter became known as the eigenstate 
thermalization hypothesis (ETH). 

The interest in the problem of thermalization, 
and in the dynamics of isolated quantum systems far from 
equilibrium in general, was recently boosted by experiments with ultracold atoms in optical 
lattices. In particular, the antagonistic results obtained with a bosonic gas in
quasi-one-dimensional geometries, where thermalization was inferred to occur in one experiment 
\cite{hofferberth07} but was not observed in another one \cite{kinoshita06}, motivated various 
theoretical studies of nonintegrable one-dimensional (1D) quantum systems after a quench.
A special property of the 1D systems that have been analyzed is the possibility to reach 
both integrable and nonintegrable regimes by adjusting parameters of the Hamiltonian. It was 
verified that close to the integrable point, ETH ceases to be valid and thermalization does 
not happen \cite{rigol09STATa,rigol09STATb}. But the absence of thermalization has also been 
linked to other factors, such as the effects of particle statistics in finite systems
\cite{rigol09STATb} and the proximity of the energy of the initial state to the energy 
of the ground state of the system after the quench \cite{rigol09STATa,rigol09STATb,RouxARXIV}.

A close inspection of the static properties of the models being studied can anticipate the 
results for the dynamics~\cite{Santos2010PRE}. In finite 1D lattices, the 
integrable-chaos transition for fermions has been shown to require larger integrability-breaking 
terms than for bosons, which can explain the lack of thermalization of the former in certain
regimes~\cite{Santos2010PRE}. Models describing realistic systems involve only few-body 
interactions, therefore random matrices need to be substituted by banded matrices 
\cite{French1970,Brody1981,IzrailevREVIEW,Flores2001,Kota2001}. In clean systems, such as the ones involved
in recent studies \cite{rigol09STATa,rigol09STATb,Santos2010PRE,RigolARXIV}, these sparse
matrices do not even contain random elements. Full random matrices and banded matrices 
may show similar spectral statistics, but they differ in terms of level density, the first 
showing a semicircular spectrum and the latter a Gaussian spectrum, and in terms of 
eigenstates. Contrary to full random matrices, where all eigenstates are
random vectors, in the case of finite range interactions, chaos develops only away
from the edges of the spectrum, so it is only there that the eigenstates can satisfy ETH.
This explains why nonequilibrium initial states with energy close to the 
borders of the spectrum are not expected to thermalize~\cite{Santos2010PRE}.

Further factors that have been associated with the absence of thermalization in finite 
1D systems are the opening of a gap as one crosses a superfluid to insulator transition
\cite{Kollath2007}, and the existence of ``rare'' states, which for nonintegrable systems
have been speculated to persist in the thermodynamic limit \cite{BiroliARXIV}. In general,
the question of thermalization as one crosses a superfluid (metal) to insulator transition 
has attracted a lot of attention  
\cite{Kollath2007,Manmana2007,roux09,RouxARXIV,BiroliARXIV,RigolARXIV}. We
have recently argued that thermalization does happen in the gapped side of the
phase diagram, and that as one increases the system size it occurs deeper into
that side \cite{RigolARXIV}. We did not find evidence of the existence of rare states
in those systems \cite{RigolARXIV}.
Thermalization ceased to occur only when the system approached the atomic limit 
and the eigenstates started to localize in the momentum basis.

In the present work, we further analyze the issue of thermalization in systems that approach
a localization regime close to the atomic limit. As in Ref.~\cite{RigolARXIV}, localization 
refers here to the broad notion of contraction of the eigenstates in a particular basis set, 
instead of the more specific concept of spatial localization due to disorder. 
Disorder is absent in the systems that we consider. In comparison to Ref.~\cite{RigolARXIV}, 
an extra complication is added to our studies: when analyzing the observables of interest, 
some discrete symmetries are not removed. We then address the role of such symmetries in their 
static and dynamical properties. It is well known that the mixing of symmetries may conceal key 
features of the chaotic regime, such as level repulsion \cite{Kudo2005,Santos2009JMP}. Could it 
affect also the validity of the ETH? We show that the structure of eigenvectors that belong to 
different subspaces remain very similar in the chaotic region. As a result, EEVs do not fluctuate 
and the ETH continues to be valid. 

We focus on 1D systems of hard-core bosons (HCBs) at half-filling, with nearest neighbor (NN) hopping ($t$)
and interaction ($V$) and with next-nearest neighbor (NNN) interaction ($V'$). In the absence of 
NNN interaction, the model is integrable; we study how integrability is broken by $V'$. 
In addition, when $V'\gg t,V$ a transition to localization in the momentum basis starts to take 
place and is reflected in the inverse participation ratio (IPR) of the eigenstates of the Hamiltonian. 
We study how this second transition affects the EEVs of various few-body observables and their off-diagonal 
matrix elements. EEVs and the off-diagonal elements of the observables are related to the occurrence of 
thermalization and to the time evolution of the system after a quench, respectively.

The paper is organized as follows. Section \ref{Sec:model} describes the model
Hamiltonian studied and its symmetries. Section \ref{Sec:chaos} analyzes the 
integrable-chaos transition based on various chaos indicators. The eigenstate expectation 
values of different observables and the comparison with the microcanonical averages 
are shown in Sec.\ \ref{Sec:observables}. Section \ref{offdiag} is devoted to studying 
the behavior of the off-diagonal elements of few-body observables in the eigenstates of the 
Hamiltonian. Concluding remarks are presented in Sec.\ \ref{Sec:remarks}. Further 
illustrations about delocalization measures and observables are provided in the Appendix.

\section{System Model}
\label{Sec:model}

As mentioned in the introduction, we study a 1D HCB model with 
NN hopping $t$ and interaction $V$, and NNN interaction $V'$. The Hamiltonian is given by
{\setlength\arraycolsep{0.5pt}
\begin{eqnarray}
&&\hat{H}_{b}=\sum_{i=1}^L \left\lbrace -t\left( \hat{b}^\dagger_i \hat{b}_{i+1} 
+ \textrm{H.c.} \right) \right. \label{Eq:hamiltonian} \\
&&+V\left.\left( \hat{n}_i-\dfrac{1}{2}\right)\left( \hat{n}_{i+1}-\dfrac{1}{2}\right) 
 +V'\left( \hat{n}_i-\dfrac{1}{2}\right)\left( \hat{n}_{i+2}-\dfrac{1}{2}\right)\right\rbrace. 
\nonumber
\end{eqnarray}
}where $L$ is the size of the chain, $\hat{b}_i$ ($\hat{b}_i^{\dagger}$) is the bosonic 
annihilation (creation) operator on site $i$ and $\hat{n}_i= \hat{b}_i^{\dagger} \hat{b}_i$ 
is the boson local density operator. Hard-core bosons are not allowed to occupy the same 
site, so $b_i^2=b_i^{\dagger 2}=0$. 

Hamiltonian~(\ref{Eq:hamiltonian}) conserves the total number of particles $N_b$ and is translational 
invariant. It consists of independent blocks, where each one is associated with a value of $N_b$ and
a total momentum $k$. Here we study chains with an even number of sites and at half-filling, 
$N_{b}=L/2$, and consider all values of $k$, from 0 to $L/2$. At half-filling, other symmetries are 
found: particle-hole exists for all $k$'s and parity appears only for $k=0,L/2$. We perform full exact 
diagonalization of each $k$-sector separately for chains of 18, 20 and 22 sites. The dimension $D_k$ 
of each $k$-sector is given in Table \ref{table:dimensions}. 
Notice that we need to take into account also the double multiplicity of 
the eigenstates belonging to $k=1, \ldots L/2-1$.
The largest total Hilbert space considered 
has dimension $D=705\;432$. 

\begin{table}[h]
\caption{Dimension of $k$-sectors}
\begin{center}
\begin{tabular}{|c|c|c|c|c|}
\hline 
$L=18$ & $k=0,9 $ & $k=1,2,4,5,7,8 $ & $k=3,6$ &  \\
$D_k$ & 2704 & 2700 & 2703 & \\
\hline
$L=20$ & $k=0,10 $ & $k=1,3,7,9$ & $k=2,4,6,8$ & $k=5$ \\
$D_k$  & 9252 & 9225  & 9250 & 9226 \\
\hline
$L=22$ & $k=0,11 $ & other $k$'s & &  \\
$D_k$ & 32066 & 32065 & & \\
\hline 
\end{tabular}
\end{center}
\label{table:dimensions}
\end{table}

In what follows, $t=1$ ($\hbar=1$) sets the energy scale and the interactions 
are repulsive, $V, V' >0$. We fix $V=1$ and vary $V'$ from 0 to 10. The system is 
integrable when $V'=0$, while the addition of NNN interaction may lead to the onset 
of chaos. Moreover, there is a critical value of the NNN interaction, $V'_c=2$, below 
which the ground state is a gapless superfluid and above which it becomes a gapped 
insulator \cite{zhuravlev97}. A small bond-ordered phase develops around $V'_c=2$
\cite{schmit04}.
When $V'\gg t,V$, the system approaches the atomic limit. Due to the translational 
invariance of model (\ref{Eq:hamiltonian}), the eigenstates of the Hamiltonian
approach the eigenstates of the total momentum operator.

\section{Quantum Chaos Indicators}
\label{Sec:chaos}

Notions of phase-space trajectory and Lyapunov exponent, which are used to distinguish 
regular from chaotic motion in classical mechanics, have no meaning in the quantum domain. 
Nevertheless, criteria exist to separate quantum systems whose classical counterpart are 
chaotic from those whose classical counterpart are regular. Signatures of quantum chaos 
are obtained from the eigenvalues and from the eigenvectors of the Hamiltonian.

\subsection{Spectral observables}

Spectral observables, such as level spacing distribution, level number variance, and 
spectral rigidity are intrinsic indicators of the integrable-chaos transition 
\cite{HaakeBook,Guhr1998,ReichlBook}. However, a main disadvantage associated with the 
computation of these quantities is the need to identify and separate all symmetry sectors 
of the system. It is only after the separation that the spectrum may be unfolded and 
the analysis carried out.

The distribution of spacings $s$ of neighboring energy levels is the most frequently used 
observable to study short-range fluctuations in the spectrum
\cite{HaakeBook,Guhr1998,ReichlBook}. Quantum levels of  integrable systems are not 
prohibited from crossing and the distribution is Poissonian, 
\begin{equation}
P_{P}(s) = \exp(-s).
\end{equation}
In non-integrable systems, crossings are avoided and the level spacing distribution is 
given by the Wigner-Dyson distribution, as predicted by RMT. The form of the Wigner-Dyson 
distribution depends on the symmetry properties of the Hamiltonian. Ensembles of random 
matrices with time reversal invariance and rotational symmetry, the GOEs, lead to 
\begin{equation}
P_{WD}(s) = \frac{\pi s}{2}\exp\left( -\frac{\pi s^2}{4}\right).
\end{equation}
The same distribution form is expected for Hamiltonian~(\ref{Eq:hamiltonian}) in the 
chaotic regime, even though $\hat{H}_{b}$ has only two-body interactions and does not 
contain random elements.

\begin{figure}[!htb]
\centerline{\includegraphics[width=0.475\textwidth]{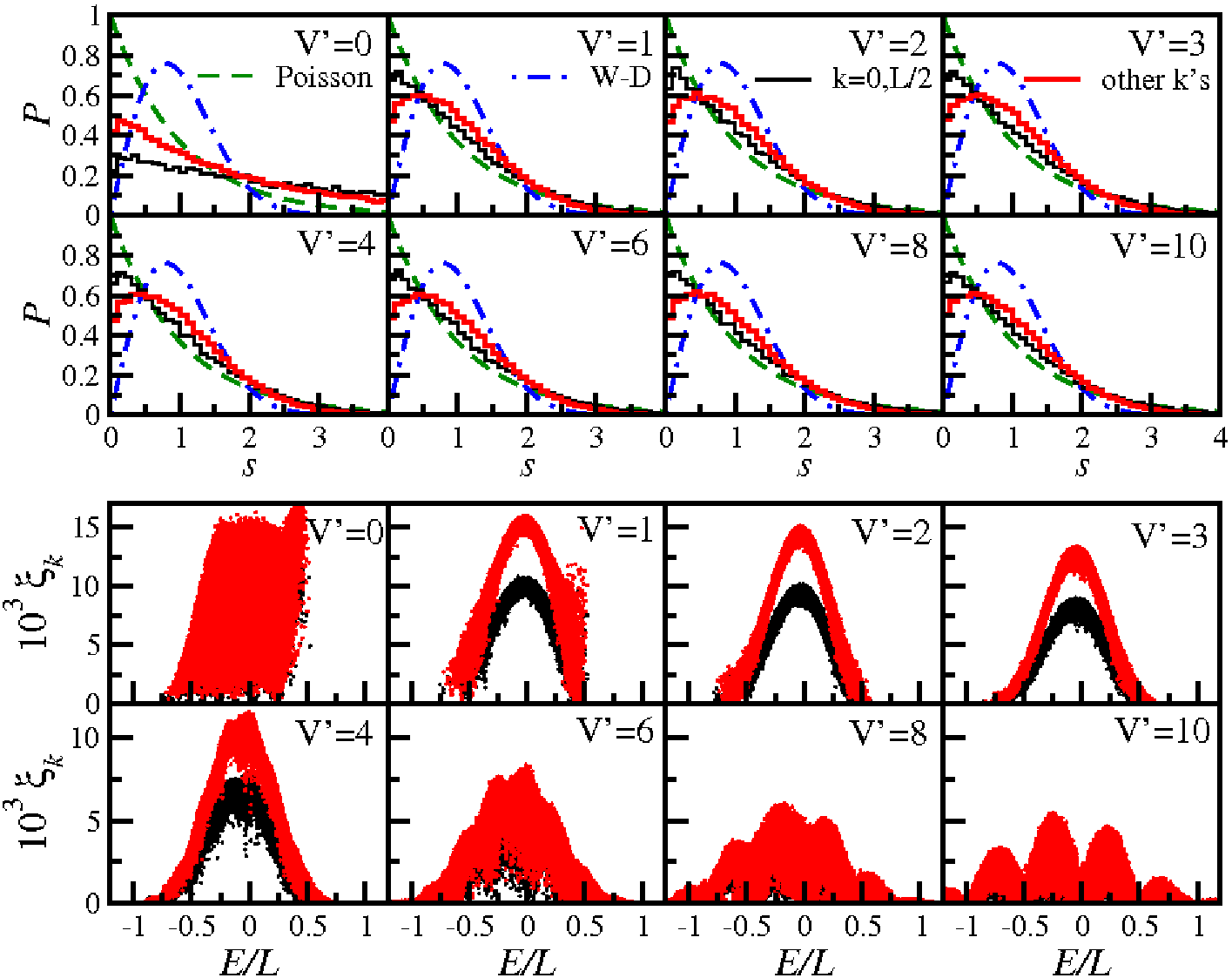}}
\caption{(Color online.)  Top panels: Level spacing distributions. Results are shown for
the average between sectors $k=0$ and $k=L/2$ [black solid line], the average between all other $k$-sectors
[light (red) solid line], 
the Poisson distribution, and the Wigner-Dyson (W-D) distribution. 
Bottom panels: Inverse participation ratio in momentum 
space for sectors $k=0$ and $k=L/2$ [black dots] and all other $k$-sectors [light (red) dots]. 
In all cases the system size is $L=22$.}
\label{fig:PsIPR}
\end{figure}

The top panels of Fig.~\ref{fig:PsIPR} depict the level spacing distributions for 
Hamiltonian~(\ref{Eq:hamiltonian}) when $L=22$. $P(s)$ is computed for each $k$-sector 
separately and the results are then averaged between $k=0$ and $k=L/2$, and between the rest
of the $k$-sectors. The decision to perform two different averages is made because 
inside each $k$-sector, particle-hole symmetry is present for all $k$'s, but 
parity exists only for $k=0,L/2$; in addition, all the sectors in each average behave very 
similarly. The top panels in Fig.~\ref{fig:PsIPR} clearly show that the distribution of
spacings never becomes equal to $P_{WD}(s)$, instead they show an intermediate behavior 
between $P_{P}(s)$ and $P_{WD}(s)$, even though we do expect these systems to be chaotic away from $V'=0$.
This occurs because we have not separated the subspaces 
according to {\em all} symmetries. The mixing of the remaining symmetries 
in each $k$-sector obscures 
the effects of level repulsion~\cite{noteBohigas}. Notice that, as expected from the 
amount of remaining symmetries, 
sectors $k=0$ and $k=L/2$ are the ones further away from a $P_{WD}(s)$ distribution.
The overall behavior of $P(s)$ in these systems is certainly in contrast with the results 
presented in Refs.\ \cite{Santos2010PRE,RigolARXIV} for other 1D chains, and shows that  
the presence of as many as one discrete symmetry may hinder the signatures of quantum chaos.

\subsection{Delocalization measures}

Quantities that focus on the eigenvectors, as delocalization 
measures~\cite{Izrailev1990,ZelevinskyRep1996}, are not intrinsic indicators of the 
integrable-chaos transition, since they depend on the basis in which the computations 
are performed. But, contrary to spectral observables, we show here that these quantities 
do not necessarily require a separated analysis of each symmetry sector.

The degree of delocalization of individual eigenvectors may be measured, for example, 
with the inverse participation ratio (IPR), denoted here by
$\xi$, or the information (Shannon) entropy S
\cite{Izrailev1990,ZelevinskyRep1996}. The former is also sometimes referred to as 
number of principal components (NPC). For an eigenstate $|\Psi_{\alpha}\rangle$ of 
Hamiltonian (\ref{Eq:hamiltonian}) written in the basis vectors $|\phi_j\rangle$ as 
$|\Psi_{\alpha}\rangle = \sum_{j=1}^{D_{k}} c^j_{\alpha} |\phi_j\rangle$, IPR and S 
are respectively given by
\begin{equation}
\xi_{\alpha} \equiv \frac{1}{\sum_{j=1}^{D_k} |c^j_{\alpha}|^4}
\label{IPR}
\end{equation}
and
\begin{equation}
\mbox{S}_{\alpha} \equiv -\sum_{j=1}^{D_k}  |c^j_{\alpha}|^2 \ln |c^j_{\alpha}|^2.
\label{entropy}
\end{equation}
The above quantities measure the number of basis vectors that contribute to 
each eigenstate, that is, how spread each state is in the chosen basis.

The choice of basis is usually determined by the information one is after and by 
possible computational limitations. Here, we consider the momentum basis, given that
for large values of $V'$ the system exhibits localization in $k$-space. This is an 
interesting effect that results from approaching the atomic limit while imposing 
translational symmetry. Other relevant bases include the mean-field 
basis~\cite{ZelevinskyRep1996}, which corresponds to choosing the eigenstates of 
the integrable Hamiltonian ($V'=0$) as a basis and therefore captures localization 
as $V'\rightarrow 0$ \cite{Santos2010PRE,RigolARXIV}, and the
site basis, which is meaningful in studies of spatial localization.

GOEs lead to extreme delocalization, their eigenvectors are random vectors
where the amplitudes $c_j^{\alpha}$ are independent random numbers.
The average over the ensemble gives $\mbox{S}^{\text{GOE}}\sim\ln(0.48 D_k)$ 
and $\xi^{\text{GOE}}\sim D_k/3$ \cite{Izrailev1990,ZelevinskyRep1996}. 
Since Hamiltonian~(\ref{Eq:hamiltonian}) has only two-body interactions,
the eigenstates of our system in the chaotic limit may approach the GOE result
only away from the edges of the spectrum~\cite{Brody1981,Kaplan2000,Kota2001}.

The bottom panels of Fig.~\ref{fig:PsIPR} show IPR in the $k$-basis ($\xi_k$) 
for {\em all} $k$-sectors for the same values of $V'$ used in the study of the 
level spacing distributions (top panels). While the results for $P(s)$ hardly change 
with $V'$, three different regimes can be singled out from the behavior of $\xi_k$.
(i) When $V' \rightarrow 0$, the values of $\xi_k$ fluctuate considerably for 
states very close in energy, which agrees with our expectations for a system in the 
integrable regime~\cite{Santos2010PRE}. (ii) For intermediate values of $V'$ 
[$0<V'\lesssim 5$ for $L=22$], $\xi_k$ becomes a smooth function of energy, 
which suggests the crossover to chaos~\cite{Santos2010PRE}. (iii) At large values 
of $V'$, energy bands are created and the $\xi_k$ once again fluctuates considerably. 
We also notice that in the scenario (iii), $\xi_k$ decreases significantly, signaling
localization in $k$-space. The two transitions~\cite{RigolARXIV}, from integrability 
to chaos as $V'$ increases from zero and from chaos to localization in $k$-space as 
$V'\rightarrow \infty $, are therefore clearly captured by $\xi_k$, despite 
the inclusion of eigenstates from different subspaces.

Two separated curves are clearly distinguished in the bottom panels of 
Fig.~\ref{fig:PsIPR} when $V'=1,2,3,4$. This is caused by two combined factors. 
First, particle-hole symmetry exists for all $k$-sectors, but parity is only 
present for $k=0$ and $k=L/2$. Thus, the eigenstates from the two latter subspaces 
cannot spread as much as those pertaining to $k=1,2, \ldots L/2-1$ and so
exhibit smaller values of $\xi_k$. Second, the structures of the eigenstates 
from different $k$-sectors, containing the same number of internal symmetries, 
are very similar and do not fluctuate in the chaotic region. As a result, 
the two domains, regular with large fluctuations and chaotic with no fluctuations, 
are well distinguished. This explains why the analysis of $\xi_k$ is an efficient 
way to detect the transition to chaos even when we do not separate the eigenstates 
according to all of their symmetry sectors. 

With increasing system size, we find that
chaotic behavior beyond $V'_c$ can be observed for larger values of $V'$. This is
shown in the Appendix (A.1), where we argue that similar conclusions, as the ones
presented here with $\xi_k$, are reached with the Shannon entropy.

\subsection{Structural entropy}

Further information about the structure of the eigenvectors may be obtained with 
the so-called structural entropy, which is defined as \cite{Pipek1992,Jacquod2002}

\begin{equation}
\mbox{S}_{\text{str}}  \equiv \mbox{S} - \ln \xi.
\label{Sstructural}
\end{equation}
$\mbox{S}_{\text{str}}$ contains the contribution to the information entropy which 
is not found in the IPR. It is an attempt to better distinguish states that may have 
similar levels of delocalization, but different structures. In the case of a GOE, 
the states are uniform and $\mbox{S}_{\text{str}}^{\text{GOE}} \approx 0.3646$.

\begin{figure}[!htb]
\centerline{\includegraphics[width=0.475\textwidth]{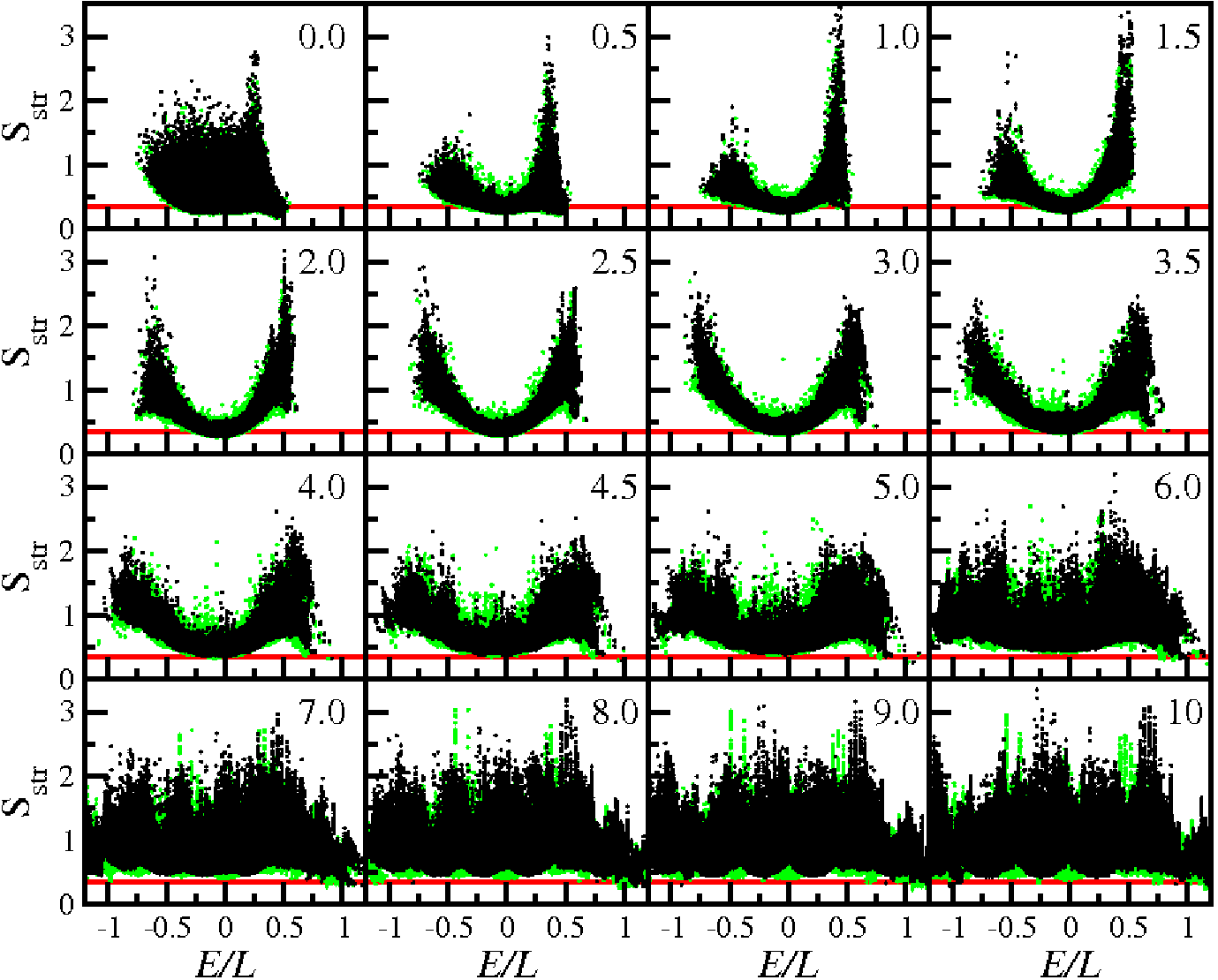}}
\caption{(Color online.) Structural entropy in the momentum basis vs energy per site
for all $k$-sectors. 
Black dots: $L=22$, light (green) dots: $L=20$. The values of $V'$ are indicated in the panels.
The solid line correspond to  $\mbox{S}_{\text{str}}^{\text{GOE}} \approx 0.3646$.}
\label{fig:structural}
\end{figure}

Since S$_{\text{str}}$ does not aim at measuring the actual extension of the
eigenstates, but instead at capturing their structures, vectors belonging to 
different symmetry sectors  may be analyzed on a par with each other, even when 
they have different levels of delocalization. In Fig.~\ref{fig:structural} we show 
S$_{\text{str}}$ for {\it all} $k$-sectors and for two system sizes, $L=20$ and $L=22$. 
The results for all sectors are remarkably similar and superpose each other. The plots 
reveal again the same three regimes identified in Fig.~\ref{fig:PsIPR}. (i) For 
$V'\rightarrow 0$, eigenstates very close in energy have different levels of complexity, 
as typical of integrable systems. (ii) The structures of the states become comparable to 
random vectors in the middle of the spectrum when chaos is reached 
[when $0<V'\lesssim 5$ for $L=22$]. The two-body interactions are responsible for the 
bowl-shaped curve and the fluctuations at the edges of the spectrum.  (iii) As the system 
moves to localization in the $k$-basis, for large $V'$, energy bands accompanied by 
large fluctuations of the values of S$_{\text{str}}$ become evident. The fact that the 
analysis of S$_{\text{str}}$ does not require the identification and separation of 
symmetry sectors supports our claim that quantities associated with the eigenvectors 
may, in many instances, be better suited than spectral observables for studying the 
integrable-chaos transition, especially when unknown symmetries may be present.

In Fig.~\ref{fig:structural}, we present results for two system sizes, $L=20$ and $L=22$. 
The results are very similar, but S$_{\text{str}}$ for $L=20$ exhibits larger fluctuations,
particularly in the chaotic region, i.e. fluctuations decrease in the chaotic regime as $L$ 
increases. Also, for large values of $V'$, where eigenstates are grouped in bands with similar 
energies, S$_{\text{str}}$ shows that those bands shift as the system size increases.

\begin{figure}[!htb]
\centerline{\includegraphics[width=0.39\textwidth]{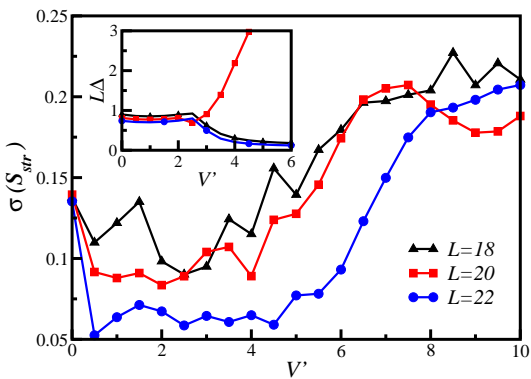}}
\caption{(Color online.) Standard deviation of S$_{\text{str}}$ vs $V'$ for eigenstates in 
the middle of the spectrum with energies varying from -5 to 5. The inset shows the gap 
(between the lowest energy states) times $L$ vs $V'$. Results are shown for lattices with
$L=18$, $L=20$, and $L=22$ sites.}
\label{fig:stDEV}
\end{figure}

In Fig.~\ref{fig:stDEV} we compare the standard deviation of the structural entropy,
\begin{equation}
\sigma(\mbox{S}_{\text{str}}) = 
\sqrt{\langle \mbox{S}_{\text{str}}^2  \rangle - 
\langle \mbox{S}_{\text{str}} \rangle^2},  
\end{equation}
for states in the middle of the spectrum for different lattice sizes. As $V'$ increases 
from zero, $\sigma $ captures the two transitions of model~(\ref{Eq:hamiltonian}). This
is particularly visible for $L=22$, where large fluctuations appear in the integrable 
domain ($V' \rightarrow 0$) and in the localization regime ($V' \rightarrow \infty$), 
while small fluctuations are associated with the onset of chaos.

In relation to the low energy behavior of these systems, the opening of the gap $\Delta$ 
between the ground state and the first excited state, signaling the onset of the superfluid to 
insulator transition (the ground state becomes four-fold degenerate in the insulating side) 
is well illustrated by the curve for $L=20$ in the inset of Fig.~\ref{fig:stDEV} [boundary 
effects conceal the transition for $L=18$ and 22]. By comparing the value of $V'$ for the 
chaos-localization transition with the value for the superfluid-insulator transition,
it becomes clear that an overlap between chaotic regime and gapped phase exists for the 
finite systems considered here.
In the case of $L=20$, for instance, the opening of the gap is already evident when $V'\sim3$
(see inset of Fig.~\ref{fig:stDEV}), while the formation of energy bands followed by the 
localization in the momentum basis requires 
$V'>4$ (see Figs.~\ref{fig:structural}, \ref{fig:stDEV}, and \ref{fig:EntL20}).

The dispersion in the main panel of Fig.~\ref{fig:stDEV} makes evident also the dependence 
of the results on $L$. Larger systems imply smaller fluctuations. Moreover, as $L$ increases, 
smaller values of $V'$ already lead to the first transition from integrability to chaos and 
larger values of $V'$ are required for the second transition from chaos to localization. 
The shift of $V'$ to larger values for the second transition shows that, as the system increases, 
chaoticity appears deeper into the gapped phase. In the thermodynamic limit, one may even 
speculate that any $V' \neq 0$ might suffice to guarantee the chaoticity of the system.
These results reinforce the claim that the superfluid-insulator transition does not affect 
the behavior of the bulk of the eigenstate of the Hamiltonian~\cite{RigolARXIV}.

The uniformization of the eigenvectors in the chaotic regime has been
manifested in our studies of IPR, S$_{\text{str}}$ and S in Figs.~\ref{fig:PsIPR},
\ref{fig:structural}, \ref{fig:stDEV} \ref{fig:Shannon}, and \ref{fig:EntL20}. The results prompt us to
advocate, as in Ref.~\cite{Srednicki1994} and references therein, 
that in certain situations quantities to measure the complexity of the eigenstates may be
better indicators of quantum  chaos than spectral observables. The analysis of the eigenstates hints 
also on what to expect in terms of thermalization.
Thermalization has since long been associated with chaos and ergodicity. At the classical level, 
the idea is well established \cite{Ford1970}, while in the quantum domain the connection is based 
on a hypothesis, the ETH.
According to the ETH~\cite{Srednicki1994},  the
eigenstate expectation values (EEVs) of few-body 
observables do not fluctuate between eigenstates that are close in energy and 
hence they coincide with the microcanonical average.  
This reflects the fact that in the chaotic regime
the structure of the eigenstates 
in a small interval of energy  may be thought as equivalent.
The smooth behavior of EEVs with energy, which is achieved in the chaotic domain, is 
discussed and illustrated in the next section.

\section{Few-Body Observables}
\label{Sec:observables}

Here, we provide numerical support for the connection between the ETH and quantum chaos. 
This is done based on the analysis of the EEVs of four different observables:

(i) the kinetic energy, 
\begin{equation}
\hat{K}=-t\sum_i \left( \hat{b}^\dagger_i \hat{b}_{i+1} + \textrm{H.c.} \right),
\label{KE}
\end{equation} 

(ii) the interaction energy, 
\begin{eqnarray}
\hat{I}&=&V\sum_i \left( \hat{n}_i-\dfrac{1}{2}\right)\left( \hat{n}_{i+1}-\dfrac{1}{2}\right)\nonumber\\
&&+V'\sum_i\left( \hat{n}_i-\dfrac{1}{2}\right)\left( \hat{n}_{i+2}-\dfrac{1}{2}\right),
\label{IE}
\end{eqnarray} 

(iii) the momentum distribution function,
\begin{equation}
\hat{n}(k)=\frac{1}{L}\sum_{i,j} e^{-k(i-j)} \hat{b}^\dagger_i \hat{b}_{j},
\label{dist_n}
\end{equation} 

(iv) and the density-density correlation structure factor,
\begin{equation}
\hat{N}(k)=\frac{1}{L}\sum_{i,j} e^{-k(i-j)} \hat{n}_i \hat{n}_j.
\label{Nk}
\end{equation}
Since the operator for the total number of bosons commutes with the Hamiltonian,
the expectation value of $\hat{N}(k=0)$ is simply $\langle \hat{N}(k=0)\rangle=N^2_b/L$.
This value is set to zero in what follows.
$\hat{K}$ and $\hat{n}(k)$ are one-body observables, local and non-local, respectively; while
$\hat{I}$ and $\hat{N}(k)$ are two-body observables, local and non-local, respectively.
$\hat{K}$ and $\hat{n}(k)$ are routinely measured in cold gases experiments.

\begin{figure}[!htb]
\centerline{\includegraphics[width=0.475\textwidth]{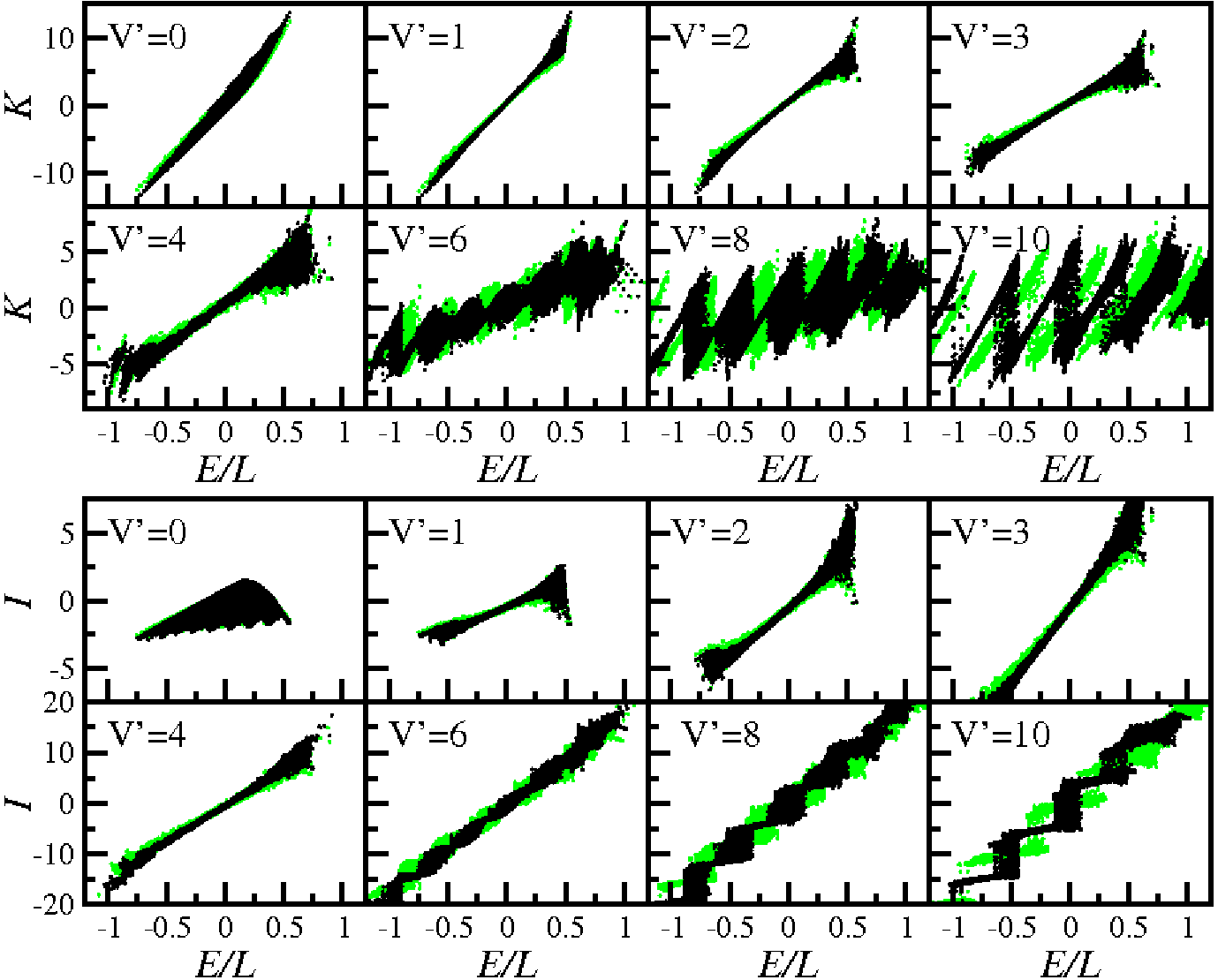}}
\caption{(Color online.) Eigenstate expectation values of $\hat{K}$ and $\hat{I}$ vs energy 
per site for 
the full spectrum, including all momentum sectors. Black dots: $L=22$, light (green) dots: $L=20$.}
\label{fig:KE_IE}
\end{figure}

\begin{figure}[!htb]
\centerline{\includegraphics[width=0.475\textwidth]{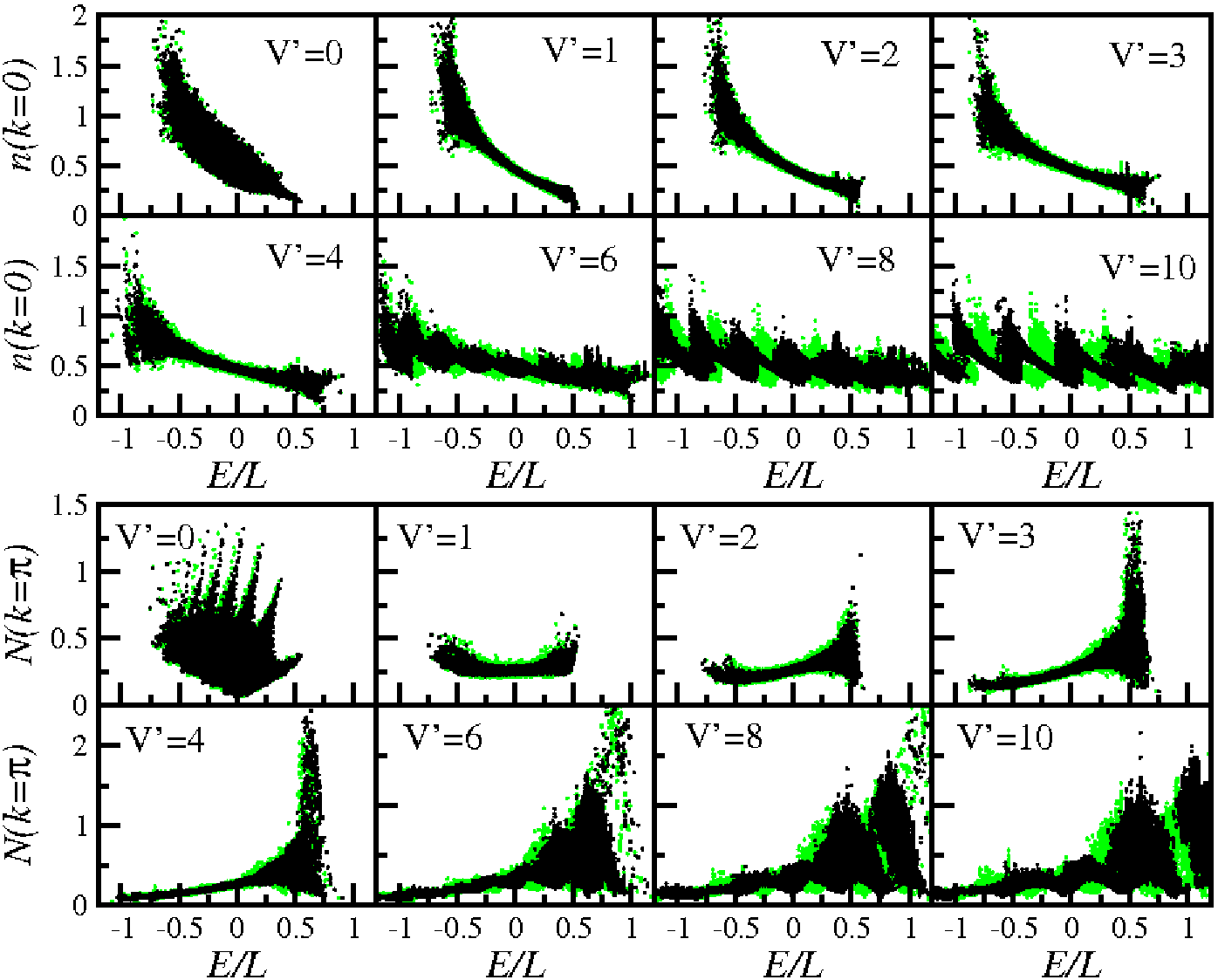}}
\caption{(Color online.) Eigenstate expectation values of $\hat{n}(k)$ and $\hat{N}(k)$ vs energy 
per site for 
the full spectrum, including all momentum sectors. Black dots: $L=22$, light (green) dots: $L=20$.}
\label{fig:n0Nk}
\end{figure}

Figures~\ref{fig:KE_IE} and \ref{fig:n0Nk} show the EEVs for the four observables defined above. 
The results parallel the findings for the eigenstates obtained in the previous section.
As $V'$ increases and one departs from integrability, the fluctuations are significantly
reduced away from the borders of the spectrum, and ETH becomes valid. 
It is remarkable that despite the inclusion of EEVs for all disconnected $k$-sectors, the 
results are still very similar for eigenstates that are close 
in energy \cite{rigol09STATa,rigol09STATb}. Contrary to the eigenstates, 
which showed at least a difference in the level of delocalization depending on their $k$-sector, 
states for $k=0,L/2$ being less spread than
for the other $k$'s (cf. bottom of Fig.~\ref{fig:PsIPR}), the behavior of the 
EEVs in all $k$-sectors is very similar.
Hence, it is not surprising that if
some discrete symmetries are not accounted for when diagonalizing the Hamiltonian,  
ETH will still be valid in the chaotic regime. No separation of the EEVs occurs for different symmetry sectors. 

The smooth behavior of EEVs with energy, which is characteristic of the chaotic domain, 
continues to hold beyond the superfluid-insulator transition (compare Figs.~\ref{fig:KE_IE} 
and \ref{fig:n0Nk} with the inset of Fig.~\ref{fig:stDEV}), further confirming that the 
latter is irrelevant for the discussion of the validity of ETH.
By increasing $V'$ even further, the eigenstates finally 
begin to localize in $k$-space and large 
variations of the EEVs for states close in energy reappear. In this limit, the separation 
of the expectation values into energy bands becomes evident. 

The comparison between the results for different system sizes in Figs~\ref{fig:KE_IE} and 
\ref{fig:n0Nk}, $L=20$ and $L=22$, shows that (i) as the system size increases, the fluctuations 
between EEVs of states that are close in energy decrease in the chaotic region, and
(ii) in the localization regime, which is reached for large $V'$, the position of the energy bands
for $L=20$ and $L=22$ do not coincide. The comparison also reinforces the disconnection between 
the behavior of low energy states and the bulk of states. As seen in the inset of Fig.~\ref{fig:stDEV}, 
a gap opened for $L=20$, but boundary effects prevented it in $L=22$. This difference has no 
consequences in the results for EEVs, which are very similar for both $L$'s. 

\subsection{Eigenstate thermalization hypothesis}

Strong evidence of the validity of the ETH is 
established once EEVs are seen to be very similar between eigenstates
that are close in energy. This, in turn, implies that thermal averages and the EEVs will be also
very similar. The analysis above shows that this should occur in the chaotic regime.
To quantify this statement for finite systems, we compute the deviation of the EEVs for an observable $O$
with respect to the microcanonical result ($\Delta^\textrm{mic}$), defined as
\begin{equation}\label{devmic}
\Delta^\textrm{mic} O \equiv \frac{\sum_{\alpha}\,
|O_{\alpha \alpha}-O_\textrm{mic}|}{\sum_{\alpha}\,O_{\alpha \alpha}}.
\end{equation}
In Eq.\ \eqref{devmic}, the sum runs over the microcanonical window, $O_{\alpha\alpha}$ are the 
EEVs of the operator $\hat{O}$, and the microcanonical expectation values $O_\textrm{mic}$ are obtained 
from
\[
O_\textrm{mic} = \frac{1}{{\cal N}_{E,\Delta E}} 
\sum_{\underset{|E-E_{\alpha}|<\Delta E}{\alpha}} O_{\alpha\alpha},
\] 
where ${\cal N}_{E,\Delta E}$ is the number of energy eigenstates with energy in the 
window $[E-\Delta E, E+\Delta E]$. In what follows, we will also refer to $\Delta^\textrm{mic}O $ as the 
average fluctuations of the EEVs.

\begin{figure}[!htb]
\centerline{\includegraphics[width=0.475\textwidth]{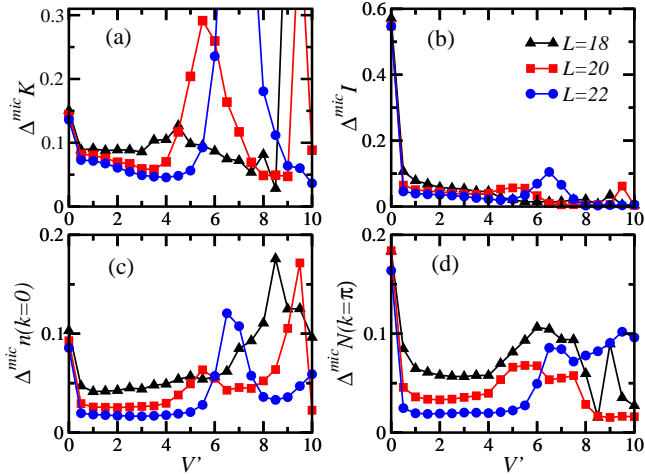}}
\caption{(Color online.) Average relative deviation of eigenstate expectation values
with respect to the microcanonical result as a function of $V'$.  The average is performed 
over all eigenstates (including all momentum sectors) that lie within the window 
$[E-\Delta E,E+\Delta E]$, with $\Delta E=0.4$. Results are shown for lattices with $L=18$, 
$L=20$, and $L=22$ sites. The effective temperature is $T=5$, which for $L=22$ corresponds 
to $E=-2.9680$ for $V'=0$, $E=-4.3126$ for $V'=2$,  $E=-7.7462$ for $V'=4$, $E=-13.1636$ 
for $V'=6$, $E=-20.4001$ for $V'=8$, and $E=-29.2205$ for $V'=10$.}
\label{fig:MICROvsL}
\end{figure}

Figure~\ref{fig:MICROvsL} shows the relative deviations $\Delta^\textrm{mic}K$, $\Delta^\textrm{mic}I$, 
$\Delta^\textrm{mic}n(k=0)$, and $\Delta^\textrm{mic}N(k=\pi)$ averaged over all momentum sectors and for 
all eigenstates that lie within a window $[E-\Delta E, E+\Delta E]$, where $\Delta E=0.4$.
While the results should not depend on the exact value of $\Delta E$ around a reasonable choice, 
the selection becomes subtle for large values of $V'$, where energy bands are formed and 
the number of states for small energy windows decay significantly. Our choice was made to
guarantee that all $k$-sectors, for all system sizes and for all values of $V'$,
have a sufficiently large number of eigenstates in $[E-\Delta E, E+\Delta E]$. 
[For more discussion of the dependence of the results on $\Delta E$, see the Appendix (A.2)]. 
The value of $E$ is selected according to the effective temperature $T$ that we chose to 
study. Performing the analysis in terms of a single temperature allows for a fair comparison
of all systems sizes and values of $V'$. The effective temperature, $T_{\alpha}$ of an 
eigenstate $|\Psi_{\alpha}\rangle$ with energy $E_{\alpha}$ is defined as
\[
E_{\alpha} = \frac{1}{Z} \mbox{Tr} \left\lbrace \hat{H} e^{-\hat{H}/T_{\alpha}} \right\rbrace ,
\]
where
\[
Z=\mbox{Tr} \left\lbrace e^{-\hat{H}/T_{\alpha}} \right\rbrace .
\]
Above, $\hat{H}$ is Hamiltonian (\ref{Eq:hamiltonian}), $Z$ is the partition function with 
Boltzmann constant $k_B=1$, and the trace is performed over the full spectrum. 

As seen in Fig.~\ref{fig:MICROvsL}, the average fluctuations of the EEVs for all observables considered 
decrease as $V'$ increases from zero and the integrable-chaos transition takes place, which goes along with
the validity of the ETH in the chaotic domain. The dependence of the results on system size 
is also clear: the average fluctuations decrease for larger systems. In addition, the width of the 
interval of values of $V'$ for which the EEVs approach the thermal averages can, in general, be seen to
increase with $L$, which brings the validity of ETH deeper into the gapped phase. On the other hand,
beyond the chaotic domain, as $V' \rightarrow \infty$ and the system approaches
the atomic limit, large fluctuations reappear.
As the system starts to localize in the momentum basis and energy bands are formed 
(when $V' \gtrsim 6$ for $L=22$), $\Delta^\textrm{mic}K$ and $\Delta^\textrm{mic}n(k=0)$ become even larger 
than in the integrable regime [cf. panels (a) and (c)].
This is a consequence of the large fluctuations that occur especially for the kinetic energy and 
the momentum distribution function (see Figs.~\ref{fig:KE_IE} and \ref{fig:n0Nk}) in the windows 
of eigenstates associated with the chosen effective temperature $T=5$.

As discussed in Ref.~\cite{Santos2010PRE}, the proximity to the ground state
prevents thermalization in systems with few-body interactions, even in the chaotic domain, 
since chaos does not develop at the edges of the spectrum. For the systems considered 
here, the presence of a gap is an additional hindering factor for the thermalization of nonequilibrium
initial states with low energies. This is because far from the ground state now means that the energy 
of the time-evolving state has to be greater than the energy of the first excited state, 
which is determined by the gap. This implies that the minimal effective temperature at which 
thermalization will occur increases as the gap in the system increases.

\begin{figure}[!htb]
\centerline{\includegraphics[width=0.475\textwidth]{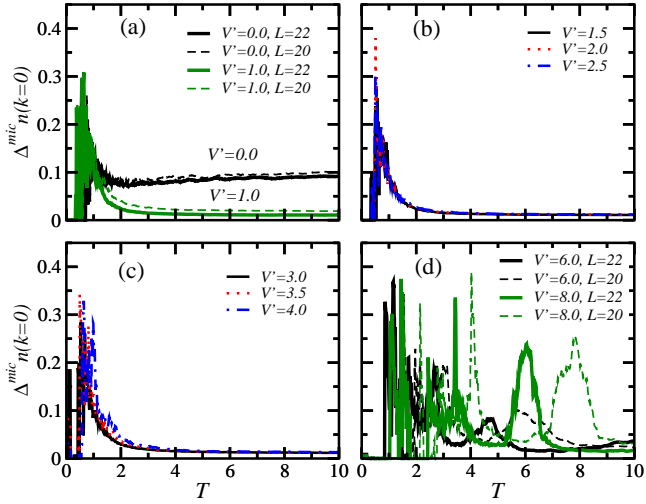}}
\caption{(Color online.) Average relative deviation of eigenstate expectation values of 
$\hat{n}(k)$ with respect to the microcanonical prediction as a function of the effective 
temperature $T$ of the eigenstates. Panels (a) and (d) depict results for $L=20$ and $L=22$.
In panels (b) and (c) only $L=22$ is shown, since the results for $L=20$ are very similar.
The average is performed over all eigenstates (including all momentum sectors) that lie within the window 
$[E-\Delta E,E+\Delta E]$, with $\Delta E=0.1$.}
\label{Fig:fluct}
\end{figure}

The study of the average fluctuations of the EEVs as a function of temperature further supports the 
conclusions above. In Fig.~\ref{Fig:fluct}, we present results for $\Delta^\textrm{mic}n(k=0)$ 
vs $T$ for ten different values of $V'$ and for temperatures $T\leq 10$. As  $V'\rightarrow 0$ 
and we approach the integrable regime, large values of $\Delta^\textrm{mic}n(k=0)$ appear for all 
temperatures considered, [cf.\ panel~(a)]. Contrary to that, in the chaotic domain, large values 
of $\Delta^\textrm{mic}n(k=0)$ are restricted to low temperatures, while at large $T$, 
$\Delta^\textrm{mic}n(k=0)$ saturates at small values [cf.\ panels~(b) and (c)]. This corroborates 
our statements that the validity of ETH goes hand in hand with the onset of chaos and holds 
away from the edges of the spectrum. Far from chaoticity, when $V' \rightarrow \infty$,
large fluctuations are seen for various temperatures, and the peaks of $\Delta^\textrm{mic}n(k=0)$, 
associated with the energy bands, move in temperature as $L$ increases. 

It has been discussed in Ref.~\cite{BiroliARXIV} that, {\it for local observables}, the deviation of the 
EEVs from the microcanonical average [given by Eq.\ \eqref{devmic}] vanishes as the system size increases.
This result is independent of whether the system is integrable or not. In our figures in this section, 
we have clearly shown that, for any given system size, the deviation of the EEVs from the microcanonical 
average is, in general, larger away from the chaotic regime, no matter whether the observable is local
or nonlocal. How those fluctuations vanish as the system size increases can depend on whether the system 
is integrable or not, and is something that deserves further investigation. In Fig.~\ref{fig:MICROvsL},
the deviations of the EEVs from the microcanonical result, in particular for the nonlocal observables 
$n(k)$ and $N(k)$, are seen to decrease faster with system size in the chaotic regime.

We should stress, however, that our calculations in the chaotic regime not only show that the average 
deviations of EEVs for all our observables decreases as one increases the system size, 
but also that the same occurs with the extremal fluctuations of the individual EEVs. This provides 
a more rigorous test of the validity of the ETH.

We have studied the normalized extremal fluctuation of an observable $O$, defined as,
\begin{equation}\label{devmic2}
\Delta^\textrm{mic}_e O \equiv \left| \frac{\max O - \min O}{O_\textrm{mic}} \right|.
\end{equation}
The maximum and minimum values of $O_{\alpha\alpha}$, $\max O$ and $\min O$,
are extracted from the same energy window $[E-\Delta E, E+\Delta E]$ used to obtain
the microcanonical expectation value. 

\begin{figure}[!htb]
\centerline{\includegraphics[width=0.475\textwidth]{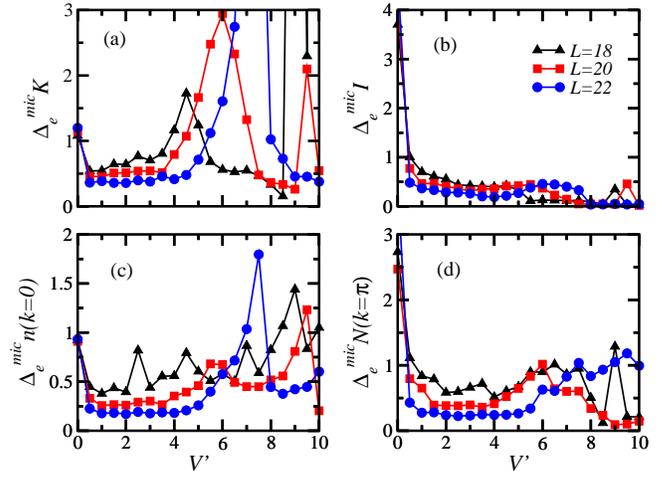}}
\caption{(Color online.) Normalized extremal fluctuations of eigenstate expectation values
as a function of $V'$.  All eigenstates from all momentum sectors that lie within the window 
$[E-\Delta E,E+\Delta E]$, with $\Delta E=0.4$ are taken into account. 
Results are shown for lattices with $L=18$, $L=20$, and $L=22$ sites. The effective temperature 
is $T=5$, which for $L=22$ corresponds to $E=-2.9680$ for $V'=0$, $E=-4.3126$ for $V'=2$, 
$E=-7.7462$ for $V'=4$, $E=-13.1636$ for $V'=6$, $E=-20.4001$ for $V'=8$, and $E=-29.2205$ for $V'=10$.}
\label{fig:MaxMinvsL}
\end{figure}

Figure~\ref{fig:MaxMinvsL} shows the normalized extremal fluctuations $\Delta_e^\textrm{mic}K$, 
$\Delta_e^\textrm{mic}I$, $\Delta_e^\textrm{mic}n(k=0)$, and $\Delta_e^\textrm{mic}N(k=\pi)$ for 
all eigenstates from all momentum sectors that lie within a window $[E-\Delta E, E+\Delta E]$, 
where $\Delta E=0.4$. This corresponds to the worst scenario, where maximum and minimum values
of $O_{\alpha\alpha}$ may belong to different $k$-sectors. Figure~\ref{fig:MaxMinvsL} mirrors some 
of the features already seen in Fig.~\ref{fig:MICROvsL}: for any given system size, as $V'$ 
increases, the extremal fluctuations for all considered observables first decrease as the integrable-chaos 
transition takes place and then increase again as the systems approaches the atomic limit. However, 
in contrast to Fig.~\ref{fig:MICROvsL}, Fig.~\ref{fig:MaxMinvsL} makes it evident that the extremal 
fluctuations of the EEVs for all observables decrease with increasing system size {\it only} 
in the chaotic region, as expected for the validity of the ETH.

\section{Predictions for the Dynamics}\label{offdiag}

In this section, we discuss what to expect for the time evolution of an arbitrary 
initial state under the unitary dynamics dictated by Hamiltonian (\ref{Eq:hamiltonian}).

For an isolated quantum system, the time evolution of an initial state 
$|\psi_{ini} \rangle$ is determined by
\[
|\psi (t) \rangle = \sum_{\alpha} C_{\alpha} e^{-i E_{\alpha} t} 
|\Psi_{\alpha}\rangle ,
\]
where $|\Psi_{\alpha}\rangle$ are the eigenstates of the Hamiltonian and 
$C_{\alpha} = \langle \Psi_{\alpha}|\psi_{ini} \rangle$.
The expectation value of an observable $\hat{O}$ at time $t$ is given by
\begin{equation}
\langle \hat{O}(t) \rangle \equiv \langle \psi (t) |\hat{O}|\psi (t) \rangle =
\sum_{\alpha \beta} C^*_{\alpha} C_{\beta} 
e^{i(E_{\alpha}-E_{\beta})t} O_{\alpha \beta},
\label{Otime}
\end{equation}
where
\[
O_{\alpha \beta}\equiv \langle \Psi_{\alpha}|\hat{O} |\Psi_{\beta}\rangle
\]
are the matrix elements of $\hat{O}$ in the energy eigenstate basis.

The infinite time average of the observable corresponds to
\begin{equation}
 \overline{\langle \hat{O}(t)\rangle}\equiv O_\textrm{diag} = 
\sum_{\alpha} |C_{\alpha}|^2 O_{\alpha \alpha},\label{diag}
\end{equation}
where ``$\textrm{diag}$'' stands for diagonal ensemble, that is an ensemble where each state 
has weight $|C_{\alpha}|^2$~\cite{rigol08STATc,rigol09STATa,rigol09STATb}.

Our results so far indicate that $O_\textrm{diag}$ will coincide with $O_\textrm{mic}$, that is 
thermalization will occur in the chaotic regime (where ETH is valid), whenever the 
initial state has an expectation value of the energy that is not close to the edges 
of the spectrum and for a distribution of $|C_{\alpha}|$ that is sufficiently narrow. 
The latter has been argued to be the case for generic quenches \cite{rigol08STATc}. 
When ETH is not valid, the outcome of $O_\textrm{diag}$ will depend on the details of the 
weights $|C_{\alpha}|^2$ and will not be, in general, in agreement with the predictions 
of standard ensembles of statistical mechanics.

The question we address here is what to expect for the time that will take
for the initial state to relax to the diagonal ensemble predictions (relaxation time) 
and for the time fluctuations that will occur about such an infinite time average. 
We might expect that longer relaxation times, as well as enhanced fluctuations 
after relaxation, should take place close to the integrable and localization regimes, 
and to edges of the spectrum in the chaotic region. These three scenarios may reduce 
the number of states $|\Psi_{\alpha}\rangle$ with a relevant role in the evolution 
of the initial state and therefore reduce the effects of dephasing in Eq.~(\ref{Otime}).
Interestingly, in previous works, the relaxation time at and close to integrability 
has not been found to be much different from the one away from integrability 
\cite{rigol09STATa,rigol09STATb,RigolARXIV}. On the other hand, the approach to 
localization in Ref.\ \cite{RigolARXIV} was shown to substantially increase the 
relaxation time and the time fluctuations after relaxation (see Fig. 3(d) in 
\cite{RigolARXIV}).
Other factors that may play a role are analyzed in what follows.

A quick relaxation to $O_\textrm{diag}$ and reduced time fluctuations require a 
nondegenerate and incommensurate spectrum, as expected for nonintegrable systems. 
However, as it has been discussed in this work, mixing of symmetries may occur even 
in the chaotic domain. In this case, states very close in energy may appear and one
may wonder if they could slow down the dephasing process in Eq.~(\ref{Otime}).

In studies of the unitary dynamics, the system is usually taken out of equilibrium
by means of a quench. One starts with an initial state of a certain Hamiltonian 
$H_{ini}$ and then instantaneously changes it to $H_{fin}$ at time $t=0$. In  
previous works~\cite{rigol09STATa,rigol09STATb,RigolARXIV}, 
$H_{ini}$ and $H_{fin}$ involved the same symmetries, and the initial state was taken
from the $k=0$ sector, which had an internal remaining symmetry, parity (these systems were 
at 1/3 filling). Surprisingly, the relaxation dynamics in the integrable and near integrable regimes,
as well as in the chaotic regime, were very similar \cite{rigol09STATa,rigol09STATb,RigolARXIV},
even though the level spacing distributions for both domains are clearly different.
Based on those results, we expect a similar behavior for the cases considered in this 
work, away from the localized regime, even if some discrete symmetries remain in the
$k$-sector where the dynamics is performed~\cite{Santos2009JMP}. This is, however, 
another subject that deserves further investigation.

From Eqs.~\eqref{Otime} and \eqref{diag}, one realizes that the time fluctuations  
of a particular observable $\hat{O}$ after relaxation can be quantified by the expression
\begin{equation}
 \langle \hat{O}(t)\rangle - \overline{\langle \hat{O}(t)\rangle } = 
\sum_{\underset{\alpha \neq \beta}{\alpha \beta}}
C^{*}_{\alpha} C_{\beta} e^{i(E_{\alpha}-E_{\beta})t} O_{\alpha \beta},
\end{equation}
which means that the off-diagonal matrix elements of the observable under consideration 
play a very important role.

\begin{figure}[!htb]
\centerline{\includegraphics[width=0.475\textwidth]{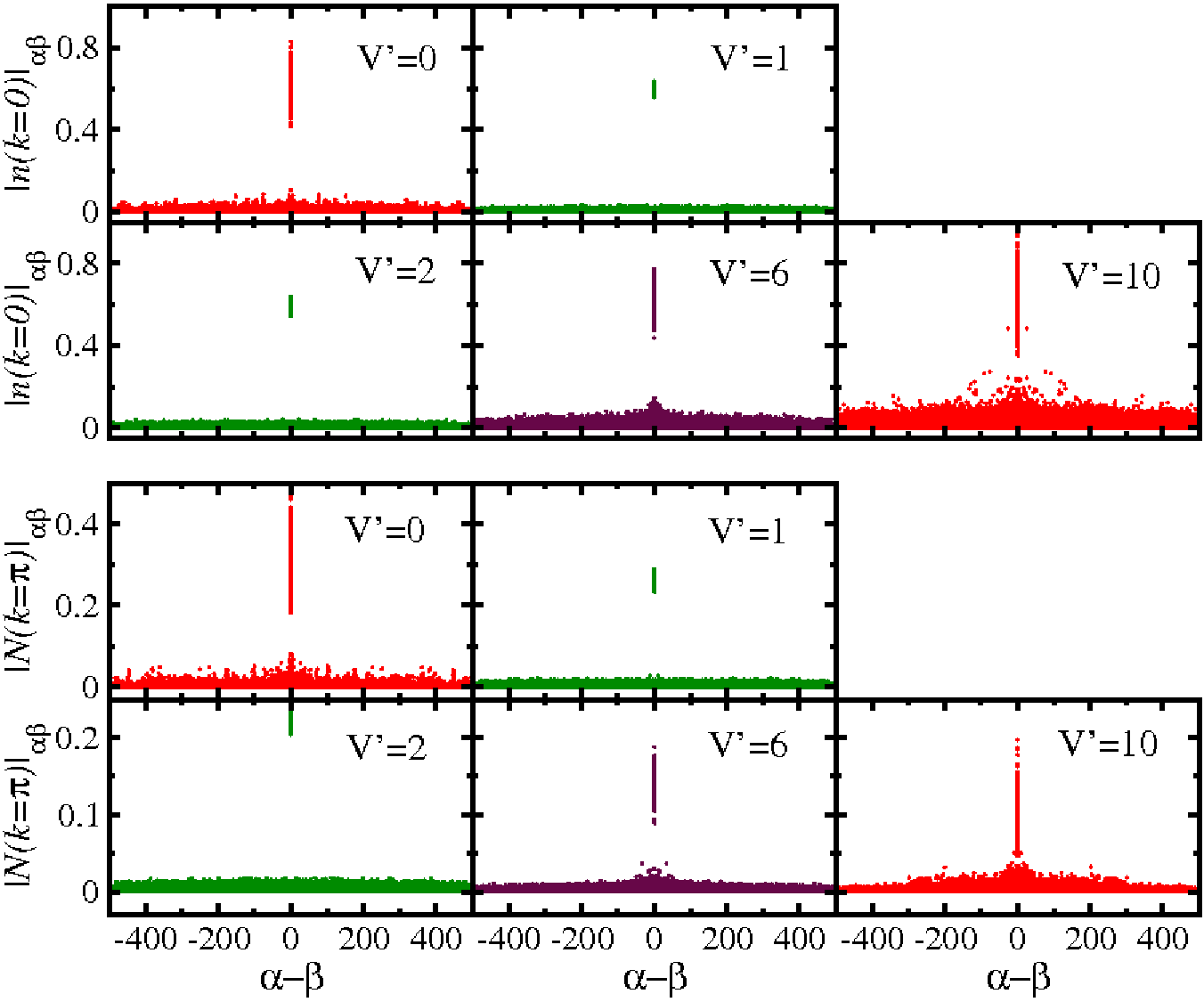}}
\caption{(Color online.) Matrix elements of $\hat{n}(k=0)$ [top panels] and 
$\hat{N}(k=\pi)$ [bottom panels]; $L=22$, $k=0$. We select as the central state the one 
that has the closest energy to the energy in the canonical ensemble corresponding to an
effective temperature $T=5.0$. For the cases depicted in the figure this corresponds 
to $E=-2.9680$ for $V'=0$, $E=-3.3748$ for $V'=1$, $E=-4.3126$ for $V'=2$, 
$E=-13.1636$ for $V'=6$, and $E=-29.2205$ for $V'=10$. A total of 500 eigenstates around 
the central one are considered; both $\alpha$ and $\beta $ run from -250 to 250.}
\label{fig:ODk0}
\end{figure}

In Fig.~\ref{fig:ODk0} we show the matrix elements for $\hat{n}(k=0)$ and 
$\hat{N}(k=\pi)$. We fix an effective temperature $T=5$ and pick the eigenstate that 
has energy closest to it as the central state and 500 states more around it. These 501
eigenstates are used to compute the matrix elements, where $\alpha$ and $\beta$ 
correspond to each one of the states, running from -250 to 250. Overall, the further 
away the element is from the diagonal, the smaller it becomes. The off-diagonal 
elements can be seen to be very small in the chaotic regime, so one expects the
time fluctuations after relaxation to be small. At integrability, $V'=0$, the 
off-diagonal matrix elements are seen to be slightly larger than in the chaotic 
regime, but it is only for large values of $V'$ where we find that the off-diagonal 
elements become very large. So, in the latter regime, as expected, time fluctuations 
after relaxation will be large. This is in agreement with the dynamics observed in 
Ref.\ \cite{RigolARXIV}.

\begin{figure}[!htb]
\centerline{\includegraphics[width=0.475\textwidth]{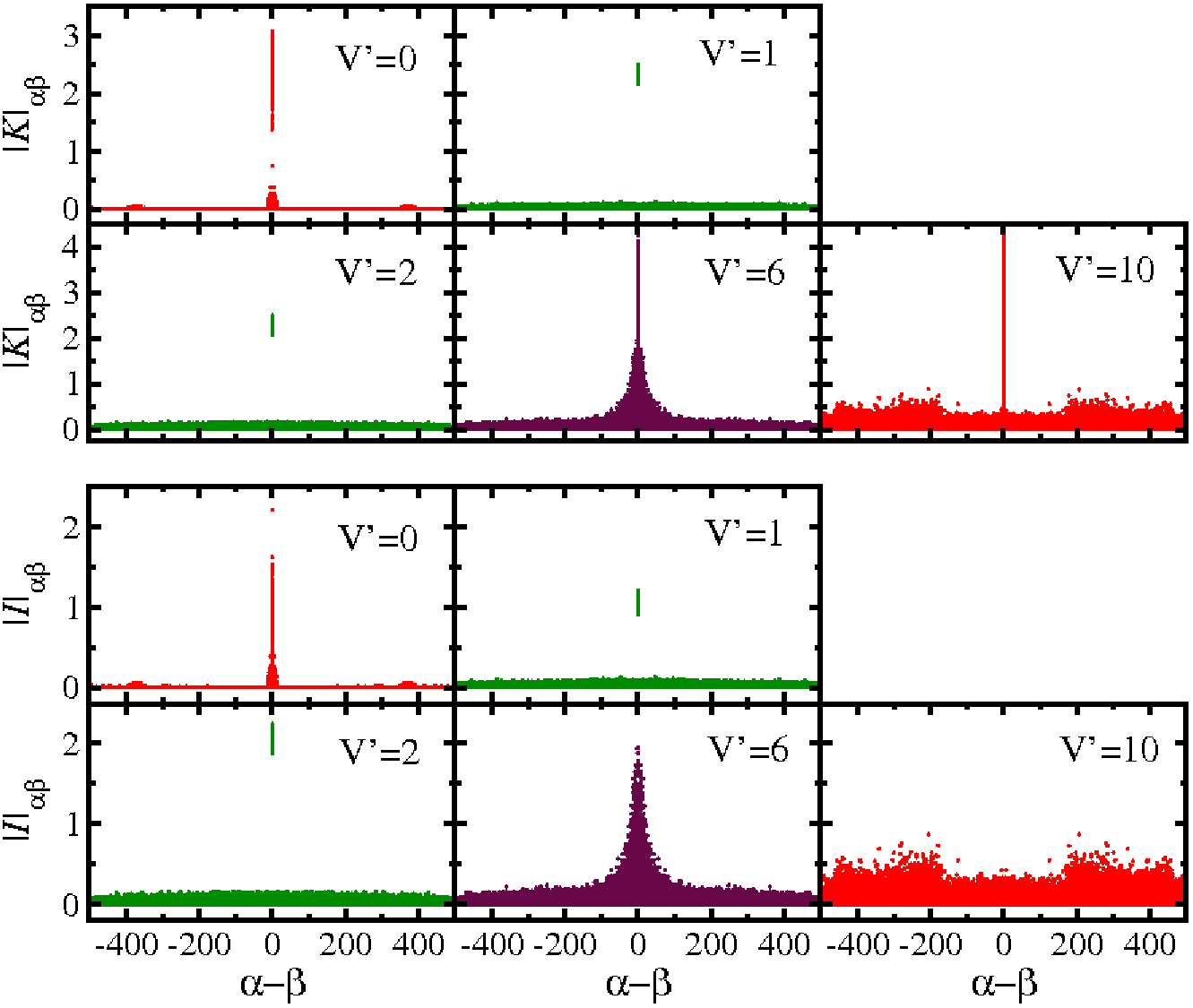}}
\caption{(Color online.) Matrix elements of $\hat{K}$ [top panels] and 
$\hat{I}$ [bottom panels]; $L=22$, $k=0$. The data correspond to the same eigenstates
of the Hamiltonian as in Fig.\ \ref{fig:ODk0}. In the bottom right panel (for $V'=10$), the
diagonal elements of $\hat{I}$ are beyond the interval presented in the plot.}
\label{fig:ODKE}
\end{figure}
 
The results for the matrix elements of $\hat{K}$ and $\hat{I}$ (Fig.~\ref{fig:ODKE}) are 
qualitatively similar as the ones for $\hat{n}(k=0)$ and $\hat{N}(k=\pi)$ (Fig.~\ref{fig:ODk0}).
However, quantitative differences are also evident and hint that the dynamics of different 
observables in experiments may exhibit quantitative differences, particularly 
when the system is approaching localization for large values of $V'$.

\section{Conclusions}
\label{Sec:remarks}

We have studied half-filled quantum chains of hard-core bosons with repulsive interactions.
The considered systems are integrable in the presence of nearest-neighbor (NN) hopping and 
interactions. The addition of next-nearest-neighbor (NNN) interactions, which are characterized 
by the parameter $V'$, leads to two different transitions: from integrability to chaos, as $V'$ 
increases from zero, and then from chaos to localization in momentum space, 
as $V' \rightarrow \infty$ and the system approaches the atomic limit. We have 
investigated the validity of the eigenstate thermalization hypothesis (ETH) in the three regimes.
ETH is found to hold whenever chaos develops. In this domain, the eigenstate expectation 
values (EEVs) for states close in energy become very similar.

Our results have confirmed previous works~\cite{Santos2010PRE,RigolARXIV} which showed that 
quantum chaos in finite systems, and thus the validity of ETH, depends on the system size $L$ 
and the range of interactions. As $L$ increases, the transition to chaos happens to smaller 
values of $V'$, but the extrapolation to the thermodynamic limit still requires further studies.
In terms of interactions, the Hamiltonian describing our system is a banded matrix, since it 
is restricted to two-body interactions. As a result, chaos develops only in the 
center of the spectrum; at the edges, the eigenstates are more localized. Initial states with 
energy close to the borders of the spectrum may therefore be unable to thermalize.

In the present work, we have focused on how the onset of chaos and the behavior of the EEVs
may be affected by two other factors: the transition to localization and the presence of symmetries.
On the way, we have shown that the opening of a gap in the ground state does not prevent 
thermalization. Even if the ground state of the system becomes an insulator, the structures of the eigenstates in the chaotic domain and 
close in energy, as well as the corresponding EEVs, do not fluctuate away from the edges 
of the spectrum. All EEVs of eigenstates close in energy were 
seen to become very similar, independently of the $k$-sector they belong to and of discrete 
symmetries that may not have been accounted for during the diagonalization. This means that 
we do not find signs of rare states~\cite{BiroliARXIV} in those systems. Moreover, the range of 
values of $V'$ over which the ETH holds increases with $L$, carrying the viability of 
thermalization deeper into the insulating phase. This corroborates our findings in 
Ref.~\cite{RigolARXIV}, where a system with 1/3 filling was considered.

In addition to the conservation of the total number of particles, the systems that we have 
analyzed presented also translational, reflection, and particle-hole symmetries. Chaos indicators 
based on the eigenvalues may miss the transition to chaos when different symmetry sectors are mixed.
Level repulsion is a main feature of chaotic systems, but to be noticeable it requires the 
examination of each symmetry sector separately. This is not the case for quantities that depend 
on the eigenvectors. In the chaotic domain, eigenstates from different symmetry sectors
were still found to have very similar structures. Therefore, the lack of fluctuations of 
quantities measuring the complexity of the eigenvectors, such as the structural entropy, or 
similarly, the lack of fluctuations of EEVs for eigenstates close in energy, has been shown to 
be a reliable alternative to identifying the chaotic region, especially in situations where 
unknown symmetries may be present.

In terms of what to expect for the dynamics, we have shown that in the chaotic regime the 
off-diagonal elements of the few-body observables of interest are very small, so that
time fluctuations after relaxation are expected to be small. At integrability, $V'=0$,
off-diagonal elements were found to be slightly larger than in the chaotic regime, but it 
was only for large values of $V'$, when the system approaches localization, that very large
off-diagonal matrix elements were seen. As expected, in the latter regime the relaxation 
dynamics will be very slow and time fluctuations after relaxation will be large.

\begin{acknowledgments}
L.F.S. thanks support from the Research Corporation. M.R. was supported by the 
US Office of Naval Research. We are grateful to Imre Varga for bringing the structural entropy 
to our attention and for motivating discussions about it. We thank Giulio Biroli, Corinna Kollath, 
and Anatoli Polkovnikov for stimulating discussions.
\end{acknowledgments}

\appendix
\label{Sec:appendix}

\section{Eigenstates and Observables}

The purpose of this appendix is to provide further illustrations for the structure of the 
eigenvectors and for the EEVS across the two transitions achieved by increasing $V'$, 
from integrability to chaos and from chaos to localization in the 
momentum basis.

\subsection{Shannon entropy}

\begin{figure}[!htb]
\centerline{\includegraphics[width=0.475\textwidth]{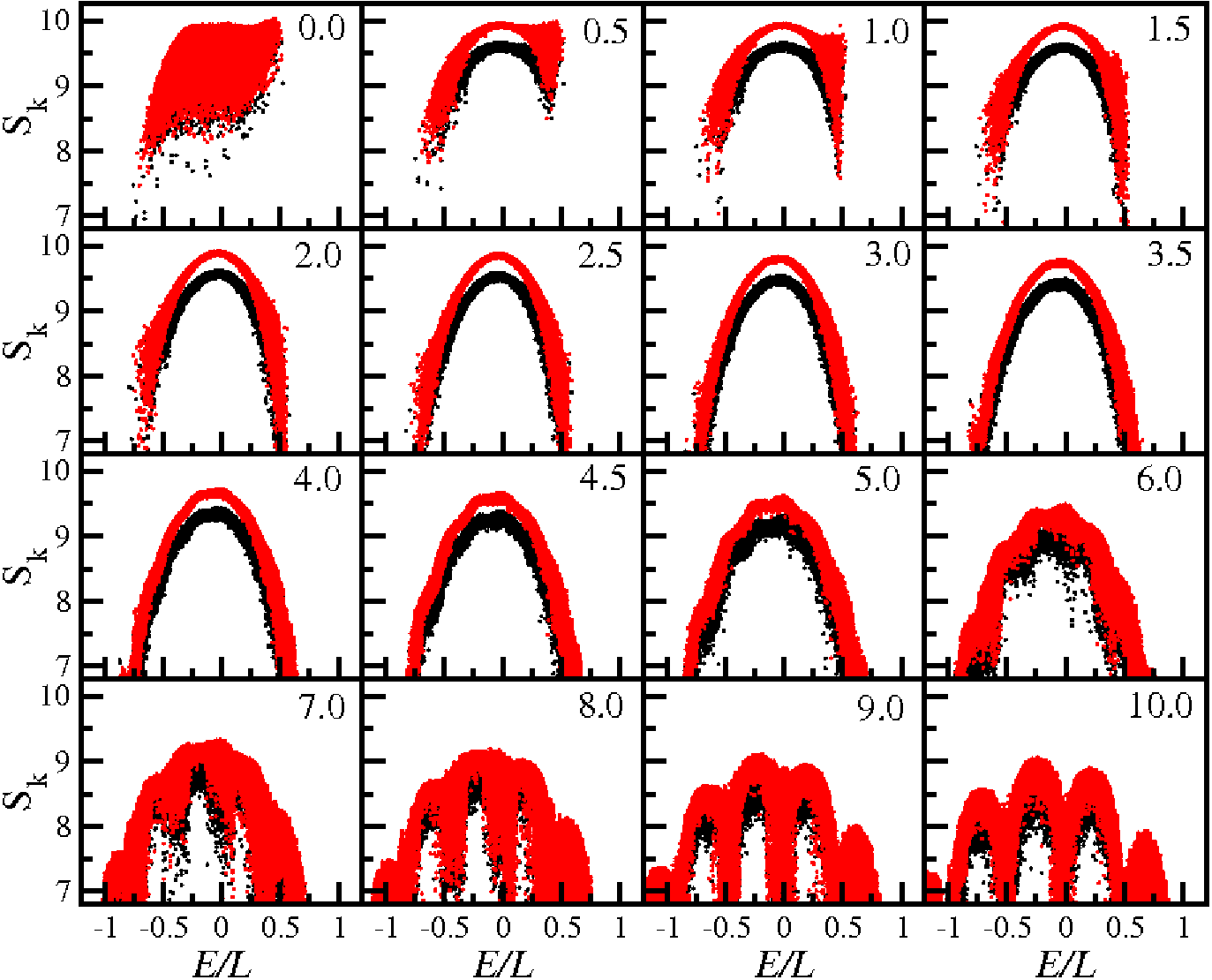}}
\caption{(Color online.) Shannon entropy in the $k$-basis vs energy per site for all $k$-sectors; $L=22$. 
Black dots: sectors $k=0$ and $k=L/2$; light (red) dots: all other $k$-sectors.}
\label{fig:Shannon}
\end{figure}

Figure~\ref{fig:Shannon} shows the Shannon entropy (\ref{entropy}) in the $k$-basis
for various values of $V'$. The results are comparable to those for IPR in Fig.~\ref{fig:PsIPR}.
S$_k$ becomes a smooth function of energy in the chaotic regime [for $L=22$, when $0<V'\lesssim 5$]. 
Here, two curves are distinguished in the middle of the spectrum. The lower curve is associated 
with the more limited capabilities for spreading of the eigenstates in sectors $k=0$ and $k=L/2$, 
where particle-hole symmetry is also present. It is remarkable, that even when all $k$-sectors 
are combined together, the Shannon entropy is still capable of identifying the chaotic region.

For $V' \rightarrow 0$, the eigenstates have very different levels of delocalization, even
when close in energy. For $V'\rightarrow \infty$, the eigenstates divide into 
energy bands and localize in the $k$-basis (small values of S$_k$). 

\begin{figure}[!htb]
\centerline{\includegraphics[width=0.475\textwidth]{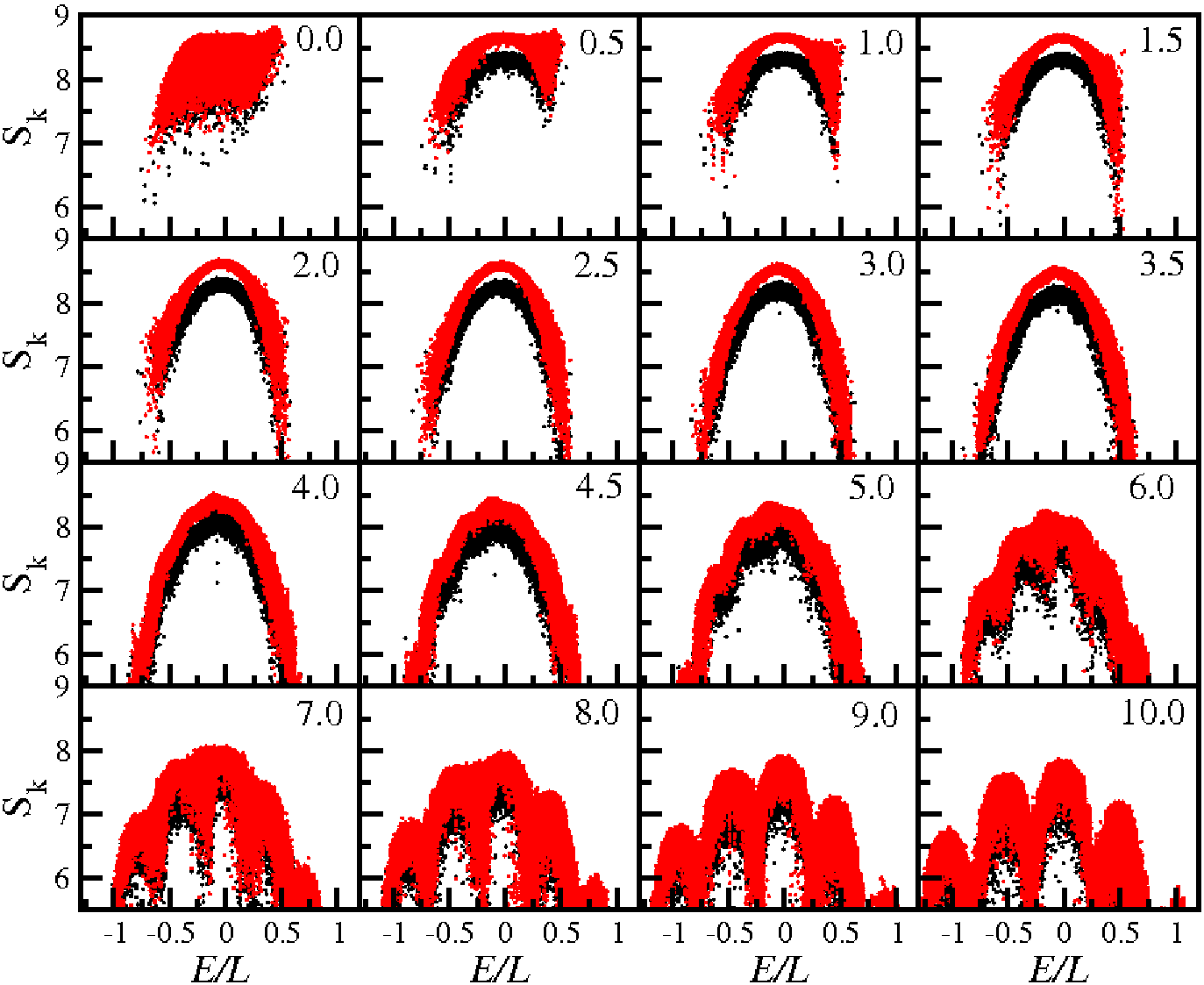}}
\caption{(Color online.) Shannon entropy in the $k$-basis vs energy for all $k$-sectors; $L=20$. 
Black dots: sectors $k=0$ and $k=L/2$; light (red) dots: all other $k$-sectors.}
\label{fig:EntL20}
\end{figure}

Figure~\ref{fig:EntL20} depicts similar results as Fig.\ \ref{fig:Shannon} but for a smaller system, 
with $L=20$. The comparison between Figs.\ \ref{fig:Shannon} and \ref{fig:EntL20} clearly shows that 
the fluctuations of S$_k$ reduce in the chaotic regime and the separation between $S_k$ for 
$k=0,L/2$ and $S_k$ for the other $k$-sectors increases with $L$. This gap may also be taken as an 
indication of the chaotic regime. With increasing system size, the chaotic regime starts at smaller 
values of $V'$ and moves towards larger values of $V'$ in the region where localization in 
$k$-space starts to become evident by the reduction of the values of S$_k$.

\subsection{Effect of $\Delta E$ in the Fluctuations of the EEVs}

\begin{figure}[!b]
\centerline{\includegraphics[width=0.475\textwidth]{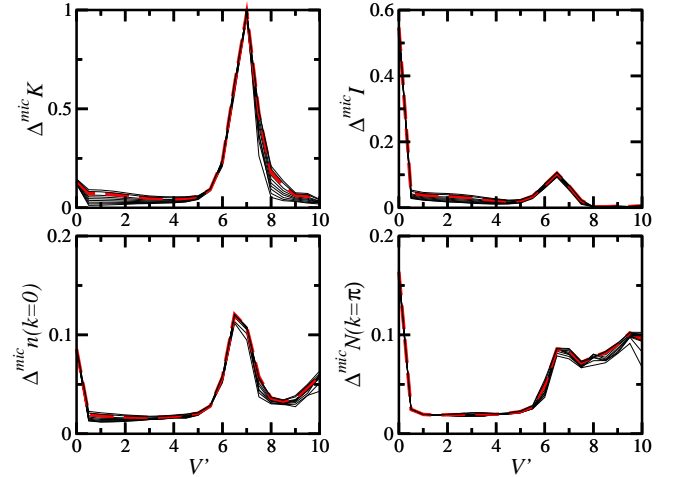}}
\caption{(Color online.) Average relative deviation of eigenstate expectation values
with respect to the microcanonical result as a function of $V'$; $L=22$. The average is 
performed over all eigenstates (including all momentum sectors) that lie within the window 
$[E-\Delta E,E+\Delta E]$. The ten curves shown in each panel are obtained for 
$\Delta E = 0.05, 0.1, 0.15, \ldots 0.5$; the thick dashed line corresponds to 
$\Delta E =0.4$, which is the value considered in Fig.~\ref{fig:MICROvsL}. The effective 
temperature is $T=5$.}
\label{fig:micro_App}
\end{figure}

Figure~\ref{fig:micro_App} shows the same results from Fig.~\ref{fig:MICROvsL} for 
$L=22$, but now for different values of $\Delta E$. The results are not much affected 
by the exact value of $\Delta E$ and the overall behavior is still the same: 
the EEVs approach the microcanonical average in the chaotic region, but show
large fluctuations in the integrable and localization regimes. 
We notice that in our plots, the larger relative fluctuations and larger effects of the window of 
energy $\Delta E$, which occur for the kinetic energy and the interaction energy in the chaotic 
regime, are related to the fact that their EEVs (see Fig.\ \ref{fig:KE_IE}), and hence their mean 
values, approach zero for the windows of eigenstates selected for our calculations.

\end{document}